\documentclass[conference]{IEEEtran}
\IEEEoverridecommandlockouts
\usepackage{cite}
\usepackage{amsmath,amssymb,amsfonts}
\usepackage{graphicx}
\usepackage{textcomp}

\usepackage{algorithm}
\usepackage{algorithmicx}
\usepackage[noend]{algpseudocode}
\usepackage[dvipsnames]{xcolor}
\usepackage[final]{changes}
\usepackage{graphicx,import}
\usepackage{amsmath}
\usepackage{siunitx}
\usepackage{amsfonts}
\usepackage{enumitem}
\usepackage{bm}
\usepackage{mathtools}

\usepackage{array}
\usepackage{booktabs}
\usepackage{array}
\usepackage{float}
\usepackage{algorithm}
\usepackage{multirow}
\usepackage{makecell}
\usepackage{comment}
\usepackage{tikz}

\usepackage{graphicx}
\definecolor{goldenrod}{HTML}{FFDF42}

\def\BibTeX{{\rm B\kern-.05em{\sc i\kern-.025em b}\kern-.08em
    T\kern-.1667em\lower.7ex\hbox{E}\kern-.125emX}}
\definecolor{linred}{rgb}{0.582, 0.039, 0.078}

\begin{document}

\title{A Proportional-Integral Model for Fractional Voltage Tripping of Distributed Energy Resources 
\thanks{This work was supported as a part of NCCR Automation, a National Centre of Competence in Research, funded by the Swiss National Science Foundation (grant number 51NF40\_225155)}
}

\author{\IEEEauthorblockN{Milo\v{s} Katani{\'c}}
\IEEEauthorblockA{\textit{Power Systems Laboratory} \\
\textit{D-ITET, ETH}\\
Zurich, Switzerland\\
mkatanic@ethz.ch}

\and
\IEEEauthorblockN{Gregor Verbi\v{c}}
\IEEEauthorblockA{\textit{Faculty of Engineering} \\
\textit{The University of Sydney}\\
Sydney, Australia \\
gregor.verbic@sydney.edu.au}
\and
\IEEEauthorblockN{John Lygeros}
\IEEEauthorblockA{\textit{Automatic Control Laboratory} \\
\textit{D-ITET, ETH}\\
Zurich, Switzerland \\
jlygeros@ethz.ch}
\and
\IEEEauthorblockN{Gabriela Hug}
\IEEEauthorblockA{\textit{Power Systems Laboratory} \\
\textit{D-ITET, ETH}\\
Zurich, Switzerland\\
hug@ethz.ch}

}

\maketitle

\begin{abstract}
 
In regions with high shares of distributed energy resources (DERs), massive disconnection of small-scale DERs in low-voltage distribution grids during disturbances poses a serious threat to power system security. However, modeling this effect in a computationally efficient way remains challenging. This paper proposes a novel proportional-integral aggregate model for predicting the fraction of tripped DERs based on the voltage at the substation connection point. The model effectively captures the cumulative behavior of the system, is simple to implement, and includes seven parameters for undervoltage tripping and seven for overvoltage tripping behavior, each with a distinct physical meaning. We further propose an optimization-based approach to tune the model parameters. Simulation results show significantly more accurate predictions compared to the DER\_A model---a standard dynamic model for aggregate DER behavior---even when the latter is optimized, with only a minor increase in model complexity.

\end{abstract}

\begin{IEEEkeywords}
Distributed Energy Resources, Voltage Ride-Through
\end{IEEEkeywords}

\section{Introduction}
\label{Intro}
\subsection{Motivation}
Power systems are experiencing the biggest and fastest change since their inception over a hundred years ago. Modern power systems are characterized by a very high number of different components. Besides legacy components, they incorporate novel power electronics interfaced components, such as electric vehicles, heat pumps, and distributed energy resources (DERs). The latter are predominantly small-scale inverter-based generation resources connected to distribution grids. 
\par
The increased share of DERs in distribution grids requires the predictability of their operational behavior at all time scales, including seconds and milliseconds. In fact, the Australian Energy Market Operator (AEMO) has observed a profound detrimental impact of distributed photovoltaics (PVs) on the power system security during large-scale disturbances due to their low voltage tripping behavior \cite{AEMO2019}. As much as 40\% of installed PVs in certain regions of the grid have been observed to disconnect from the system for a period of minutes following a disturbance endangering power system security~\cite{AEMO2019}. DER Voltage Ride-Through (VRT), which refers to the ability of DERs to remain connected to the grid during voltage deviations, is critical in mitigating such impacts. Extrapolating the current renewables installation trends in some countries, the behavior of DERs will become a crucial determining factor for power system stability and security, particularly during periods of high levels of distributed PV generation. If their behavior is not well understood and accurately reflected in power system models, it may lead to an increased conservatism in power system operation, reduced network hosting capacity, increased market costs, or in extreme cases, large-scale outages~\cite{AEMO_compliance}. Hence, it is necessary to derive accurate aggregated dynamic models of active distribution grids that are simple and flexible enough to be useful for daily power system studies.


\subsection{Literature Review on Aggregate DER Models}
The Western Electricity Coordinating Council released the DER\_A model \cite{DER_A} for aggregate DER behavior intended for transmission system stability studies. It is a reduced version of a generic model for large-scale renewable generators \cite{generic_DERA}, with approximately one-third of its parameters. The study \cite{DER_A} proposes a fractional tripping block for modeling the fraction of DERs, which disconnect due to a voltage disturbance. The proposed tripping block characteristic is too simple to accurately capture the fraction of disconnected PVs for a wide variety of operating conditions, installed systems, and grid configurations \cite{gustavo_pscc}, \cite{AEMO_PSSE}. Similar conclusions hold true for the models used in \cite{phdLVRT}. Based on \cite{DER_A}, AEMO derived the DERAEMO1 model \cite{AEMO_PSSE}. This model adapts DER\_A for application in the Australian national grid incorporating distinct voltage-tripping parameters for different grid codes under which inverters were installed in the grid. Study \cite{DG_LVRT} proposes an aggregate model for inverters VRT under large-scale disturbances, but the study assumes the knowledge of all individual inverter VRT characteristics which is not realistic for a large number of small-scale DERs. Reference \cite{gustavo_pscc} suggests a realistic VRT characteristic according to different international grid codes but assumes access to the average voltage present in the distributed grid, which is a prohibitive assumption. References \cite{JoML, Kontis2019} suggest machine learning-based, black-box methods to predict the power injection of the active distribution grids. However, the focus of these studies is not VRT, and their out-of-sample performance remains to be validated. 

\subsection{Contributions}
Given the identified research gaps, we propose a novel gray-box aggregate model for the fractional tripping behavior of DERs based on their voltage ride-through (VRT) characteristics. The model exhibits the following advantages over the existing literature:
\begin{itemize}
    \item It leverages the electrical properties of DERs and the associated grid codes and represents an intuitive way to model their aggregate disconnection rates. Therefore, its output is easily explainable, in contrast to black-box models. 
    \item It is supported by a simple optimization-based approach to fit its parameters based on synthetic data generated respecting the characteristics of employed DERs, relevant grid codes, and the parameters of the underlying distribution grid. 
    \item It achieves significantly higher accuracy compared to DER\_A, even when both models are optimized using the same parameter tuning methodology.
\end{itemize}
This study focuses on VRT capabilities and tripping logic settings of distributed PVs. We disregard the frequency-tripping logic, leaving this topic for future work. According to AEMO~\cite{AEMO_compliance}, poor voltage disturbance ride-through represents the most serious and urgent obstacle to secure and reliable operation. In addition, taking the frequency response into account is arguably less challenging because it is a global variable. In addition, we disregard the dynamics of distributed loads and model them as fixed impedances. \par

\subsection{Outline}
Section \ref{sec:DER_D} introduces the detailed DER model employed to generate data and validate the algorithm performance. Section~\ref{sec:VRT} explains the VRT settings of three inverter types used for generating detailed simulations. Section \ref{sec:DER_A} elucidates the benchmark fractional tripping block DER\_A and the proposed proportional-integral (PI) fractional tripping block. Section \ref{sec:results} evaluates the in-sample and out-of-sample performance of the modeling framework in numerical studies, and finally, Section \ref{sec:conclusion} concludes the paper.

\section{Employed DER Model}
\label{sec:DER_D}
\begin{figure}[t]
    \centering
    \includegraphics[width=\columnwidth]{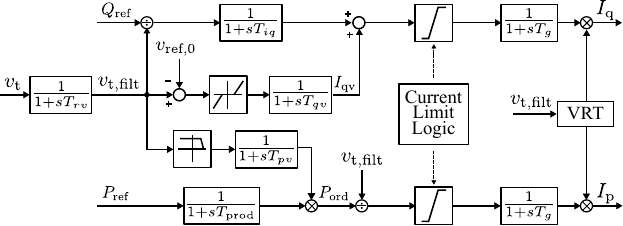}
    \caption{Block diagram of the employed detailed DER model.}
    \label{fig:pv_model}
\end{figure}

For the detailed simulation of the power systems' transient behavior in this study, we adopt the model presented in \cite{DER_A}, which complies with modern grid codes and includes complementary protection and grid-supporting functions. The block diagram of the employed model is given in Fig.~\ref{fig:pv_model}.
The detailed description of the model can be found in \cite{param_DERA}. It has been derived to represent the aggregated behavior of DERs for transmission system studies. The model assumes good tracking of the internal current control loop and neglects the fast internal dynamics (inverter switching). The actual current is assumed to follow an internal current reference with a first-order delay. The internal current reference is manipulated by the outer control functions to support voltage stability during off-nominal operating conditions. Volt--VAr and Volt--Watt responses specify varying reactive and active power output based on the locally measured voltage to protect and support the network depending on the current grid situation. \par
This study focuses on the Australian national grid and locally valid grid codes, but we believe that the findings are transferable to other power systems with high shares of renewables. The parameters of the model are set according to \cite{param_DERA} and the relevant grid codes in Australia (see Table~\ref{tab:DERD} in Appendix \ref{app:par}). The model employed for this study shown in Fig.~\ref{fig:pv_model} differs from DER\_A \cite{DER_A} as follows:
\begin{itemize}
    \item It encompasses Volt--Watt and asymmetrical Volt--VAr control loops according to AS/NZS 4777.2:2020 \cite{AS2020};
    \item It includes first-order delays for Volt--VAr and Volt--Watt control loops in accordance with the observed delayed responses during field testing of solar inverters \cite{AEMO_tests};
    \item It features detailed VRT logic for each DER consistent with the voltage measured at its terminal and the relevant grid codes in Australia (see Section \ref{sec:VRT}). The VRT block employs a binary tripping mechanism governing the on/off state of each PV unit. Further, the block does not manipulate the current references but directly the output currents because the tripping delays are explicitly considered within the VRT block (protection circuit delay is neglected). The next section dives into the details of the VRT block.
\end{itemize}

\section{Voltage Ride-Through (VRT)}
\label{sec:VRT}
VRT characteristics define the behavior of inverters during voltage disturbances usually caused by a fault. In response to voltage sags or spikes, inverters may continue to operate normally, reduce their output, cease operation, or disconnect from the grid to protect their internal components. The VRT capabilities and tripping requirements are stipulated in international standards and grid codes.\par 
Currently, three grid codes stipulate the requirements for grid connection of energy systems via inverters in Australia: AS 4777.2-2005 \cite{AS2005}, AS/NZS  4777.2:2015 \cite{AS2015}, and AS/NZS 4777.2:2020 \cite{AS2020}. Note that standards \cite{AS2005} and \cite{AS2015} are obsolete, but a significant share of PVs connected to the grid was commissioned when these standards were in force, and therefore, they still need to be considered in the operational behavior of distribution grids.\par
\begin{figure}[t]
    \centering
    \includegraphics[width=\columnwidth]{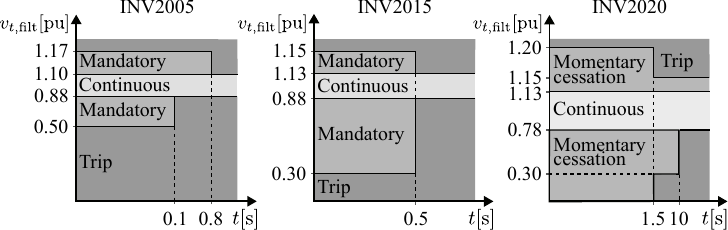}
    \caption{Voltage disturbance withstand requirements and clearing times for
trip for three modeled inverter types INV2005, INV2015, INV2020 derived from AS 4777.2-2005 \cite{AS2005}, AS/NZS 4777.2:2015 \cite{AS2015}, and AS/NZS 4777.2:2020 \cite{AS2020}, respectively. The specified values incorporate findings from the inverter bench tests \cite{AEMO_tests} and represent the average values used in the simulated fleet of inverters. Full parameter ranges are provided in Table~\ref{tab:bounds} in the Appendix.}
\label{fig:VRT}
\end{figure}

 
Grid codes specify the operating conditions of grid-connected energy systems including voltage limits for sustained operation and associated minimal delay times or maximal clearing times at which DERs shall cease to energize. \textit{Continuous} operation of each inverter refers to the nominal operating condition with full power output and without time restriction. \textit{Mandatory} operation requires an active and reactive power exchange during the specified time duration, even in the presence of a disturbance. During \textit{momentary cessation}, the inverter should stay synchronized with the grid voltage but temporarily reduce its output to zero. Once the voltage recovers to the nominal operating condition, the inverter should ramp up its power output to the pre-disturbance value. Once a DER enters the \textit{trip} region, it is disconnected, and its active and reactive power outputs remain at zero. The filtered terminal voltage $v_{t,\textrm{filt}}$ thresholds and clearing times for this study are shown in Fig.~\ref{fig:VRT}. 
In our model, the actual values used for each inverter are randomized with a distribution centered around these nominal values to reflect the diverse landscape of inverter manufacturers. The parameters are sampled from a uniform distribution to prevent infeasible settings such as negative tripping delays. The bounds of these parameter distributions are provided in Table~\ref{tab:bounds}, in the Appendix.\par
INV2005 thresholds are derived from AS 4777.2-2005 \cite{AS2005}, INV2015 from AS/NZS 4777.2:2015 \cite{AS2015}, and INV2020 from AS/NZS 4777.2:2020 \cite{AS2020}. Mandatory cessation is modeled only in INV2020 devices, as in the previous grid code versions, this behavior was not mandated. We categorize continuous and mandatory operation modes as \textit{active}, while momentary cessation and trip modes are categorized as \textit{inactive} operation modes. A change in the operating status includes an intrinsic delay necessary for the low-pass filtering of the measured terminal voltage. In addition, the transition to momentary cessation is governed by a randomized delay between 0 and \SI{100}{ms}, and the reactivation delay from momentary cessation to continuous operation is modeled with a randomized timer between 0 and \SI{400}{ms} as per \cite{AS2020}. Switch-off delays (e.g., protection circuit delay) are considered negligible. All internal disconnection timers are reset when the voltage reenters the continuous operating range. \par
Note that clearing times can range from milliseconds, meeting the needs for very fast tripping, to several seconds for situations requiring a delayed response.  As international standards specify admissible ranges of relevant parameters---giving inverter manufacturers design freedom---we take into account the results of solar inverter laboratory tests from \cite{AEMO_tests} to determine the used thresholds for our model. Additional modeling challenge represents the observed noncompliance of the solar inverter fleet with the pertinent grid codes \cite{AEMO_compliance}. The used voltage and timer bounds, nominal values, and the chosen distribution type were selected to provide a simplified yet reasonable representation of the system's behavior while maintaining manageable model complexity.


\section{Fractional Tripping Block}
\label{sec:DER_A}
To capture the tripping behavior of small-scale DERs in response to voltage disturbances without modeling each DER individually, a fractional tripping block is commonly employed. It predicts a fraction of deactivated DERs based on the substation terminal voltage and inverter tripping settings. However, accurately modeling this behavior is highly challenging due to the vast diversity in the DER landscape, pertinent grid codes, and underlying distribution grids. Moreover, the deactivation of some DERs can alter local voltage profiles, dynamically influencing the behavior of neighboring units and further complicating predictive modeling. In the following section, we present a state-of-the-art fractional tripping block introduced in \cite{DER_A}. 
\subsection{DER\_A Fractional Tripping Block}
\begin{figure}[t]
    \centering
    \tikzstyle{fblock} = [rectangle, minimum width=1cm, minimum height=3.0cm, text centered,  text width=3.5cm, draw=black]
\tikzstyle{block} = [rectangle, minimum width=1cm, minimum height=0.85cm, text centered,  text width=1.1cm, draw=black]
\def\fractional{
\tikz[remember picture,overlay]{
\draw (-1.6,-1)--(-1.4,-1) (-1.4,-1)--(-0.5,1) (-0.5,1)--(0.5,1) (0.5,1)--(1.4,-1) (1.4,-1)--(1.6,-1);
\draw[red] (-1,-0.1) -- (-0.5,0.3) (-0.5,0.3)--(0.5,0.3) (0.5,0.3)--(1,-0.1);
\draw[dash pattern= on 2pt off 4pt] (-0.5,1) -- (-0.5,-1) (0.5,1) -- (0.5,-1) (-1.4,1) -- (-1.4,-1) (-1,-0.1) -- (-1,-1);
\draw[dash pattern= on 2pt off 4pt] (-1.6,1) -- (-0.5,1);
\draw[dash pattern= on 2pt off 4pt] (-1,-0.1) -- (1,-0.1);
\draw[<->,>=stealth] (-0.2,-0.1) -- node[xshift=-0.15cm] {\small B} (-0.2,0.3);
\draw[<->,>=stealth] (0.25,-0.1) -- node[xshift=-0.15cm] {\small A} (0.25,1);
\draw[<->,>=stealth] (0,-1) -- node[xshift=-0.15cm] {\small C} (0,0.3);
}}

\begin{tikzpicture}[node distance=1.5cm]
    \node (Vtfilt) [draw=none,fill=none, xshift=-6cm] {};
    \node (Vt) [draw=none,fill=none, xshift=-6.5cm] {\small $v_{\textrm{ss,filt}}$};
    
    \node (frac) [fblock, right of=Vtfilt, xshift=0.8cm]  {};
    \node[] at ([yshift=0.2cm, xshift=0.1cm] frac) {\fractional};
    \node (vl0) [draw=none,fill=none, xshift=-5.0cm, yshift=-1.2cm] {\small $v_{l0}$};
    \node (vmin) [draw=none,fill=none, xshift=-4.48cm, yshift=-0.9cm] {\small $v_{min}$};
    \node (vl1) [draw=none,fill=none, xshift=-3.98cm, yshift=-1.2cm] {\small $v_{l1}$};
    \node (vh1) [draw=none,fill=none, xshift=-3.08cm, yshift=-1.2cm] {\small $v_{h1}$};
    \node (vh0) [draw=none,fill=none, xshift=-2.16cm, yshift=-1.2cm] {\small $v_{h0}$};
    \node (0) [draw=none,fill=none, xshift=-5.4cm, yshift=-0.8cm] {\small $0$};
    \node (1) [draw=none,fill=none, xshift=-5.4cm, yshift=1.2cm] {\small $1$};
    \node (Vrfrac) [draw=none,fill=none, above of=frac, yshift=0.6cm] {\small $v_{r,\textrm{frac}}$};
    \node (Tv) [block, right of=frac, xshift=1.4cm] {$\frac{1}{1+sT_{v}}$};
    
    \draw[thick,-] (-0.16,0) -- (0.5,0);
    
    
    \draw[thick,->,>=stealth] (0.5,0) -- (0.5,1.5);
    \draw[thick,->,>=stealth] (0.5,0) -- (0.5,-1.5);
    
    \node (Q) [draw=none,fill=none, xshift=0.4cm, yshift=1.7cm] {\small To Q control};
    \node (P) [draw=none,fill=none, xshift=0.4cm, yshift=-1.7cm] {\small To P control};
    
    \draw[thick,->,>=stealth] (Vt) -- (frac);
    \draw[thick,->,>=stealth] (Vrfrac) -- (frac);
    \draw[thick,->,>=stealth] (frac) -- (Tv);
\end{tikzpicture}
    \caption{DER\_A fractional tripping block. Adopted from \cite{gustavo_pscc}.}
    \label{fig:VRTDERA}
\end{figure}
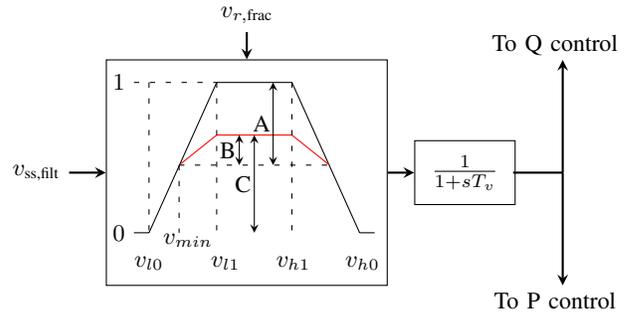

DER\_A incorporates a simple voltage-tripping logic aimed to represent the aggregated behavior of many DERs, showcased in Fig.~\ref{fig:VRTDERA}.
The idea is to approximate the aggregate voltage-tripping behavior by a linear curve between two prespecified values (upper line in Fig.~\ref{fig:VRTDERA}), each with a respective timer, intended to emulate the fraction of inactive PVs accounting for the gradient of the voltage drop along the feeder. The deactivations are determined by the filtered terminal voltage at the model's input, in this work at the substation connection point of the distribution grid $v_{\textrm{ss,filt}}$. Since only a subset of the deactivated PVs will reactivate once the voltage returns to normal operating conditions, there is another linear function with a milder gradient (lower line in Fig.~\ref{fig:VRTDERA}), depending on the parameter $V_{r,\textrm{frac}}$. The same logic applies to undervoltage and overvoltage events. The logic shown in Fig.~\ref{fig:VRTDERA} is characterized by 9 parameters: $t_{l0}, v_{l0}, t_{l1}, v_{l1}, t_{h0}, v_{h0}, v_{h1}, t_{h1}, v_{r, \textrm{frac}}$. Its main drawback is that it is too simple to accurately capture the variety of installed devices and their mutual interactions. DER systems usually encompass a range of inverter manufacturers and have been developed according to a mix of obsolete and current grid codes \cite{gustavo_pscc}. To alleviate this limitation, AEMO adopted the model for application in the Australian grid. Their extended model \cite{AEMO_PSSE} comprises three separate fractional tripping blocks according to Fig.~\ref{fig:VRT}, each with a unique set of parameters corresponding to the three relevant grid codes in Australia. Consequently, the model incorporates 27 tunable parameters. However, the findings of this study highlight a fundamental limitation of the linear fractional tripping block \cite{DER_A}. Specifically, our results show that this tripping logic fails to accurately capture the tripping behavior of many small-scale, heterogeneous DERs, and despite tuning the parameters for the specific use case, the model’s performance remains unsatisfactory. Motivated by this finding, we propose an alternative fractional tripping block offering improved accuracy. 
\par

\subsection{PI Fractional Tripping Block}

\begin{figure}[t]
    \centering
    \includegraphics[width=\columnwidth]{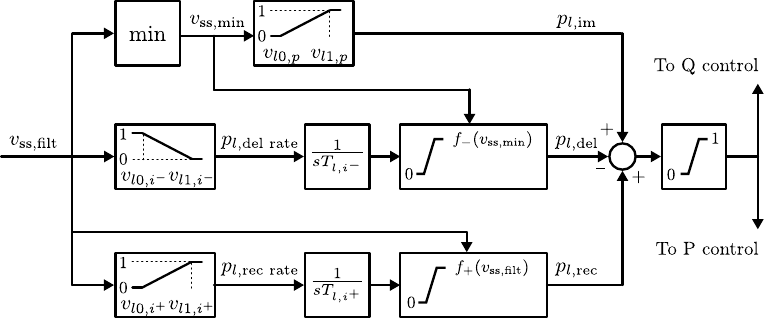}
    \caption{Proposed PI fractional tripping block for \textbf{undervoltage} (analogous logic applies for overvoltage).}
    \label{fig:PI_model}
\end{figure}

The proposed fractional tripping block for aggregate VRT of a large number of DERs is shown in Fig.~\ref{fig:PI_model}.
The output of the model denotes the fraction of active DERs. It comprises three separate branches: the proportional deactivation branch, the integral deactivation branch, and the integral reactivation branch; for this reason, we refer to it as the proportional-integral (PI) fractional tripping block. In the following, we explain the modeling framework in the case of undervoltage; the working principle for overvoltage is, however, analogous. \par
\subsubsection{Proportional deactivation branch}
\added{The proportional deactivation branch models the fraction of DERs that deactivate without delay following a voltage disturbance (except for voltage measurement filtering delay). This phenomenon occurs in the case of fast (hardware-based) deactivation within the inverter to protect the internal components. The fraction of immediately deactivated DERs is approximated as a linear function of the minimal voltage}
\begin{equation}
p_{l,\textrm{im}}(t) = \min\{\max\{\frac{v_\textrm{ss,min}(t)-v_{l0,p}}{v_{l1,p}-v_{l0,p}}, 0\}, 1\},   
\end{equation}
between two voltage values $v_{l0,p}$ $v_{l1,p}$. The output is saturated between 0 and 1. The minimal voltage is calculated as:
\begin{equation}
v_\textrm{ss,min}(t)=\min \{v_{\textrm{ss, filt}}(\tau) \; | \; \tau \in [0, t]\}.\\
\end{equation}
\subsubsection{Integral deactivation branch}
\added{The integral deactivation branch models the fraction of DERs that deactivate with a delay governed by their internal timers. Since different inverters have timer thresholds set at different values and the voltage at their terminals varies along the distribution grid feeder, the overall deactivation behavior can be complex. However, it is reasonable to assume that larger voltage deviations and longer disturbance durations lead to greater DEA deactivation. To capture this behavior, the fraction of deactivated DERs is modeled by an integrator. The disconnection rate or the integrator input is a linear function of the filtered substation voltage}:
\begin{equation}
   p_{l, \textrm{del\_rate}}(t) =  \min\{\max\{\frac{v_{l1,i^-} - v_\textrm{ss,filt}(t)}{v_{l1,i^-}-v_{l0,i^-}}, 0\}, 1\},
\end{equation}
\added{constrained between 0 and 1 by two voltage parameters $v_{l0,i^-}$ and $v_{l1,i^-}$. The integrator time constant specifies the necessary time required to deactivate 100\% of DERs. The output saturation ensures that for small disturbances, the number of disconnected DERs stays limited below a certain threshold (even if the disturbance lasts long enough). The reasoning is that due to the voltage gradient along the feeder, some DERs may measure a voltage within the continuous operating range at their terminals even when the substation voltage is out of this range. Consequently, their disconnection timers would not be activated, and they would ride through the disturbance. To capture this effect, we need to estimate the fraction of impacted DERs for any disturbance. We consider the minimal observed substation voltage. Once the minimal voltage $v_{\textrm{ss, min}}(t)$ is below $v_{l0,i^-}$, the output limit is expanded to 1; for higher minimal voltages, the output limiter is proportionally reduced. This can be described by the following formula:}
\begin{equation}
f_-(v_{\textrm{ss,min}}(t))=\min\{\max\{\frac{v_{l1,i^-} - v_\textrm{ss,min}(t)}{v_{l1,i^-}-v_{l0,i^-}}, 0\}, 1\}.
\end{equation}
\subsubsection{Integral reactivation branch}
\added{The integral reactivation branch is activated once the voltage recovers, exceeding $v_{l0,i+}$ and reentering the nominal range. The reactivation rate is also approximated by a linear function between $v_{l0,i+}$ and $v_{l1,i+}$, with the voltage span between them matching that of the integral deactivation branch:}
\begin{equation}
    p_{l, \textrm{rec\_rate}}(t) =  \min\{\max\{ \frac{ v_\textrm{ss,filt}(t) - v_{l1,i^+}}{v_{l1,i^+}-v_{l0,i^+}}, 0\}, 1\}.
\end{equation}
\added{Consequently, the parametrization of the reactivation branch is fully parametrized by a single free parameter: $v_{l0,i+}$. The fraction of reactivated DERs is again given by an integrator with a time constant $T_{l,i^+}$. The output saturation is calculated as:}
\begin{equation}
f_+(v_{\textrm{ss,filt}}(t))=\min\{\max\{ \frac{v_\textrm{ss,filt}(t) -v_{l0,i^+}  }{v_{l1,i^+}-v_{l0,i^+}}, 0\}, 1\}.
\end{equation}
Note that the integral reactivation branch in Fig.~\ref{fig:PI_model} has the opposite sign from the integral deactivation branch because of the nature of their outputs. \par 
Overall, the model comprises seven parameters each with its specific physical meaning. The pseudocode for the algorithm is illustrated in Appendix \ref{pseudocode}. To each of the relevant grid codes, we assign a PI fractional tripping block according to Fig.~\ref{fig:PI_model}. The parameters for each of the blocks can be determined based on system knowledge or data-driven approaches. In the next subsection, we propose a data-driven way to fit the parameters. Since real-life, high-sampling data capturing disturbances in distribution grids is scarce, synthetic data are generated by employing the detailed models of DERs and the underlying distribution grid.

\subsection{Optimization of Model Parameters}
The optimal set of parameters is found by an optimization algorithm such that the estimated deactivation curves match the ones obtained from the detailed simulation as closely as possible. Due to the several limiters in Fig.~\ref{fig:PI_model}, the model output is a non-smooth function of the introduced parameters. Therefore, we employ zero-order optimization methods, avoiding gradient evaluations. Specifically, we use particle swarm optimization (PSO).\par
PSO is a zero-order iterative optimization algorithm employing a set of particles called a swarm \cite{pso}. Each particle represents a potential solution of the problem and has a position in the solution space and a velocity (direction and speed of movement). Particles explore the solution space by using different movement rules while keeping track of their best position so far and the best position of their neighbors. The idea is to attract particles to move toward known high-quality solutions. Many different variants of PSO exist depending on these position update rules. If the particles stop moving (stagnation), the sequence is said to converge to a point, and the algorithm is terminated. In general, it is hard to have guarantees about the global optimality of such a solution, but, in practice, PSO often exhibits good performance.\par
The optimization problem for finding the optimal set of parameters of the proposed PI model is solved separately for the overvoltage and undervoltage parameters. The formulation for undervoltage is as follows:%
\begin{subequations}%
\begin{align}
    \label{eq:pso_obj}
    \min_{\bm{x}_l}\quad & \sum_{i=1}^{N_{l,\textrm{sim}}} \sum_{t=1}^N \left(p_{l,\textrm{PI}}(\bm{x}_l,t)-p^{(i)}_{l,\textrm{sim}}(t)\right)^2 \\
    \label{eq:pso_constr}
    \textrm{s.t.} \quad & \bm{x}_{l,\textrm{min}}\leq \bm{x}_l \leq \bm{x}_{l,\textrm{max}},
\end{align}%
\label{eq:pso}%
\end{subequations}
where subscript $l$ denotes the lower values (undervoltage), $\bm{x}_l = [v_{l0,p}, v_{l1,p}, v_{l0,i^-}, v_{l1,i^-}, T_{l,i^-}, v_{l0,i^+}, T_{l,i^+}]$ is the vector of the PI model parameters (decision variables), $N_{l,\textrm{sim}}$ represents the number of simulations in the data set, $N$ denotes the number of data points in each simulation, $p_{l,\textrm{PI}}$ denotes the output of the PI model, and $p_{l,\textrm{sim}}$ the output of the simulation with the detailed model, $\bm{x}_{l,\textrm{min}}$ and $\bm{x}_{l,\textrm{max}}$ are the bounds on physical values of the introduced parameters. For higher (overvoltage) parameters, the optimization problem is analogous, with \mbox{$\bm{x}_h = [v_{h1,p}, v_{h0,p}, v_{h1,i^-}, v_{h0,i^-}, T_{h,i^-}, v_{h0,i^+}, T_{h,i^+}]$}.\par
This formulation ensures that the model's fitness during the transient behavior is prioritized, as the simulation duration is in the range of seconds. If steady-state prediction accuracy is of greater interest, it is, however, possible to extend the simulation horizon or apply a weighted least squares approach to assign higher weights to later time steps~$t$ in the optimization problem \eqref{eq:pso_obj}--\eqref{eq:pso_constr}, thus emphasizing the model's performance at steady state. For each underlying grid code, a separate optimization problem \eqref{eq:pso} is solved.
\section{Numerical Results}
\label{sec:results}
\subsection{Detailed Model Simulation Setup}
\begin{figure}[t]
    \centering
    \includegraphics[width=0.9\columnwidth]{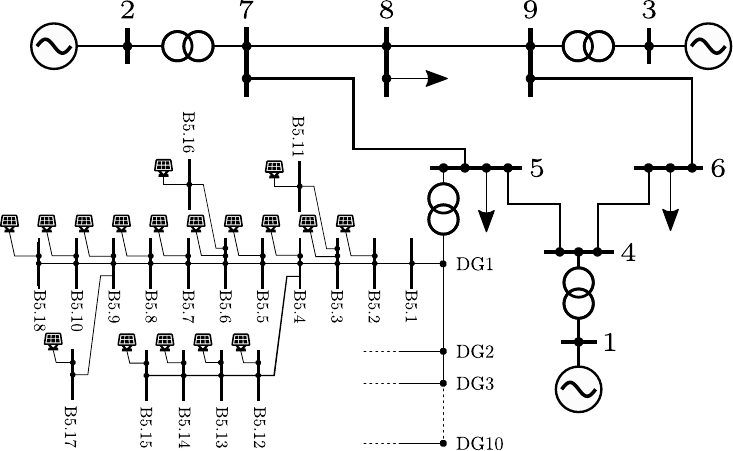}
    \caption{One-line diagram of the IEEE 9-bus and CIGRE 18-bus test cases. The simulation model comprises 10 distribution grids (DG).}
    \label{fig:IEEE9bus}
\end{figure}

We conduct transmission-distribution co-simulations on the IEEE 9-bus transmission grid \cite{9bus} and ten parallel CIGRE benchmark distribution grids \cite{cigre} connected to node $5$, as depicted in Fig.~\ref{fig:IEEE9bus}. The reason for modeling both transmission and distribution grids is their interaction, i.e., massive disconnections of DERs can significantly alter voltage profiles in the nearby transmission nodes.
PV symbols denote small-scale PVs connected to the low-voltage grid. Each PV symbol in Fig.~\ref{fig:IEEE9bus} comprises three distinct units (INV2005, INV2015, and INV2020) according to Fig.~\ref{fig:pv_model} and Fig.~\ref{fig:VRT}. The ratio of their power output is set to 0.15, 0.5, and 0.35, respectively, which is in accordance with the expected installed capacity in the Australian National Grid in 2025 \cite{AEMO_compliance}. Each of the ten parallel distribution grids represents an aggregate behavior of a large number of real low-voltage distribution grids. To represent the stochasticity in the PV generation and loading, their power outputs are sampled from a normal distribution around predefined loading and generation set points and clipped at zero. Consequently, each distribution grid experiences slightly different voltage profiles along the feeders. With 10 distribution grids, 17 PV nodes within each grid, and 3 applicable grid codes, the simulation includes a total of 510 solar inverters, each modeled with detailed voltage ride-through (VRT) characteristics as described in Section~\ref{sec:VRT}. Such a number ensures the statistical relevance of the obtained results and provides sufficient granularity to replicate the effect of tens of thousands of distributed PVs while keeping the simulation time at a reasonable level. Pre-disturbance PV production is set to \SI{34}{MW}, and the load level at \SI{17}{MW}, corresponding to the middle-of-the-day net power injection. \par  
The network is modeled in the phasor domain (algebraic equations), and the synchronous generators by the 4th-order Anderson-Fraud's model \cite{milano2010power} equipped with the IEEE DC1A automatic voltage regulator \cite{machkowski}. The effect of automatic tap changers is neglected as they are deemed too slow to impact the transient behavior in the seconds range. To direct our focus solely on DERs, we assume that all loads in the distribution grid behave according to constant impedance characteristics. It is, however, straightforward to extend our study to consider more realistic load characteristics. \par
All simulations are implemented using custom Python code as described in \cite{milano2010power} and integrated with a solver from \cite{Andersson2019} on an Intel i7-10510U CPU @ 1.80GHz with 16GB RAM. For optimization of model parameters, pyswarm python package \cite{pyswarm} is used with the default parameters. \par
\subsection{In-sample performance}
\label{subsec:insample}
\begin{figure}[t]
    \centering
    \includegraphics[]{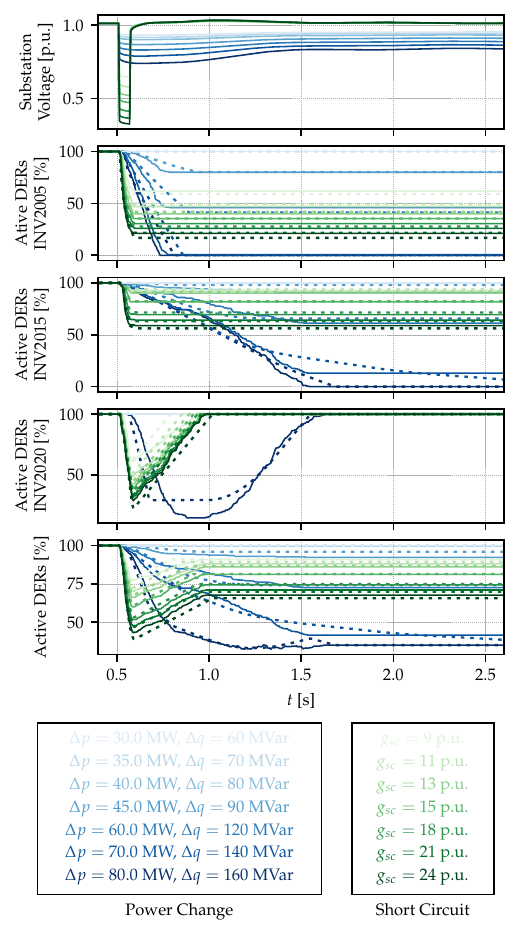}
    \caption{Voltage Ride-Through (VRT) performance of Distributed Energy Resources (DERs) under varying magnitudes of \textbf{undervoltage} disturbances. Each disturbance magnitude is indicated by a distinct color. The active DERs are shown for each of the inverter types and their weighted sum. Solid lines represent the values from the detailed model, whereas dashed lines depict the predicted values from the proposed PI aggregate model.}
    \label{fig:graph1_under}
\end{figure}

\begin{figure}[t]
    \centering
    \includegraphics[]{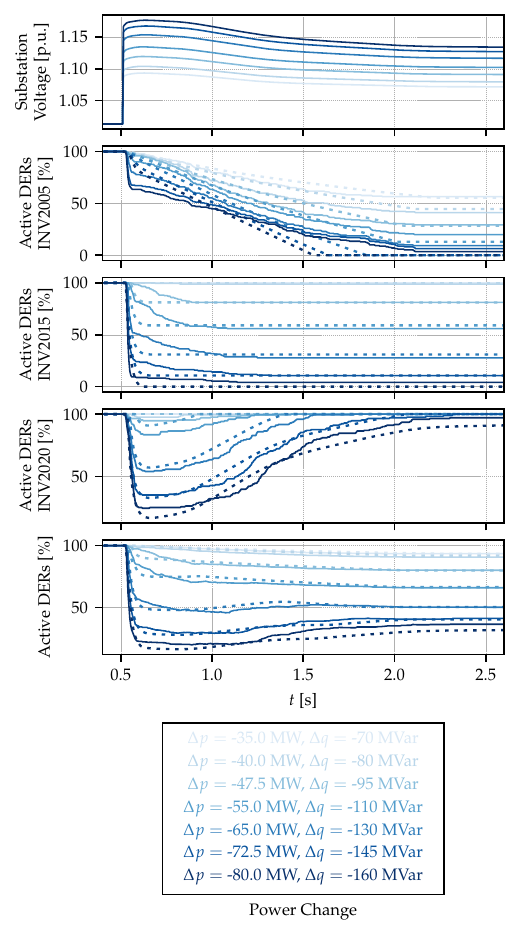}
    \caption{Voltage Ride-Through (VRT) performance of Distributed Energy Resources (DERs) under varying magnitudes of \textbf{overvoltage} disturbances. Each disturbance magnitude is indicated by a distinct color. The active DERs are shown for each of the inverter types and their weighted sum. Solid lines represent the values from the detailed model, whereas dashed lines depict the predicted values from the proposed PI aggregate model.}
    \label{fig:graph1_over}
\end{figure}

We split the simulations based on the disturbance type in overvoltage and undervoltage events. As disturbances at node 5, we run varying magnitude power injection step changes and three-phase symmetric faults with varying short circuit conductance $g_{sc}$, which is cleared after \SI{60}{ms}. The substation voltage evolutions and DER disconnection rates are shown in Fig.~\ref{fig:graph1_under} for undervoltage and Fig.~\ref{fig:graph1_over} for overvoltage disturbances. The term \textit{active DER} refers to DERs within continuous or mandatory operating regions. The figures show, from top to bottom: substation voltage, fraction of active DERs for the three inverter types, and the weighted total of active DERs. The weighted total is calculated by normalizing the fraction of inverters in each code based on their prevalence in the grid. Load changes are shown in blue and short circuits in green; greater intensity indicates more severe disturbances. The ratio of active power change to reactive power change is 1:2. We can observe that the outputs of the PI aggregate model (dashed lines) closely match the detailed simulations (solid lines). Additionally, we conclude that the underlying grid code decisively influences the overall VRT behavior. Therefore, it is beneficial to adapt models to locally valid conditions. 
The optimal parameters are reported in Table~\ref{tab:par_opt_under} and Table~\ref{tab:par_opt_over}. The optimal solution is found in less than \SI{5}{min} for all three inverter types.
\begin{table}[t]
    \caption{Optimized parameters for the \textbf{undervoltage} output of the proposed PI fractional tripping block.}
    \centering
    \setlength{\tabcolsep}{5pt}
    \begin{tabular}{c|ccccccc}
        \toprule
        \textbf{Par. }&{$v_{l0,p}$}&{$v_{l1,p}$}&{$v_{l0,i^-}$}&{$v_{l1,i^-}$} &$T_{l,i^-}$&$v_{l0,i^+}$&$T_{l,i^+}$ \\
        \midrule
        INV2005& 0.000&0.848&0.787&0.887&0.321&N/A&N/A\\
        INV2015&0.014&0.559&0.818&0.874&1.073&N/A&N/A\\
        INV2020&0.000&0.396&0.722&0.783&0.102&0.760&0.551\\
        \bottomrule
    \end{tabular}
    
    \label{tab:par_opt_under}
\end{table}
\begin{table}[t]
    \caption{Optimized parameters for the \textbf{overvoltage} output of the proposed PI fractional tripping block.}
    \centering
    \setlength{\tabcolsep}{5pt}
    \begin{tabular}{c|ccccccc}
        \toprule
        \textbf{Par. }&{$v_{h1,p}$}&{$v_{h0,p}$}&{$v_{h1,i^-}$}&{$v_{h0,i^-}$} &$T_{l,i^-}$&$v_{l0,i^+}$&$T_{l,i^+}$ \\
        \midrule
        INV2005& 1.148&1.350&1.050&1.147&1.199&N/A&N/A\\
        INV2015&1.106&1.175&1.175&1.197&0.957&N/A&N/A\\
        INV2020&1.182&1.216&1.043&1.100&0.013&1.186&0.307\\
        \bottomrule
    \end{tabular}   
    \label{tab:par_opt_over}
\end{table}
DERs compliant with the obsolete grid codes do not have the reactivation possibility (momentary cessation); therefore, the corresponding parameters are set to N/A.\par
Table~\ref{tab:error} compares the accuracy of the proposed PI model against the existing models from the literature. We employ DER\_A with parameters from \cite{param_DERA} and DERAEMO1 \cite{AEMO_PSSE}. Furthermore, to ensure a fair comparison, we also optimize the parameters of DERAEMO1 in the same manner as those for the proposed model DER\_A\_PI. The models with parameters optimized for the generated data are marked with an asterisk~(*) in Table~\ref{tab:error}.
\begin{table}[t]
    \caption{Comparison of mean absolute errors (MAE) of different models for fractional tripping of DERs (in-sample performance).}
    \centering
    \setlength{\tabcolsep}{3pt}
    \begin{tabular}{c|cccc}
        \toprule
        \textbf{MAE [\%]}&\textbf{DER\_A}&\textbf{DERAEMO1}&\textbf{DERAEMO1*}&\textbf{DER\_A\_PI*}  \\
        \midrule
        Under& 17.1&8.2&6.4&1.0\\
        Over&20.9&23.3&2.6&1.4\\
        \bottomrule
    \end{tabular}
    
    \label{tab:error}
\end{table}
The results specify the average percentage error in the fraction of deactivations across all disturbances and time steps for the weighted average of all three inverter types. Firstly, we note the importance of calibrating the parameters to the local grid codes and operating points, as models with uncalibrated parameters tend to perform relatively poorly. Furthermore, despite parameter tuning, DERAEMO1 still exhibits a significant prediction error. This is especially evident in the case of undervoltage disturbances. This fact indicates an inherent limitation in the model structure. In contrast, the proposed model outperforms others by a significant margin and achieves around 1\% error on average for both undervoltage and overvoltage disturbances.\par
\subsection{Out-of-sample performance}
\added{
To evaluate the out-of-sample performance and generalization capability of the proposed fractional tripping block, we run new simulations with different disturbances to the previous subsection. Power injection step changes are simulated with an active power change to reactive power change ratio of 1:0.8, and the short circuits are extended to last \SI{120}{ms} instead of \SI{60}{ms}. Table~\ref{tab:out_error} illustrates the performance of DERAEMO1* and DER\_A\_PI*.}
\begin{table}[t]
    \caption{Comparison of mean absolute errors (MAE) of different models for fractional tripping of DERs (out-of-sample performance).}
    \centering
    \begin{tabular}{c|cc}
        \toprule
        \textbf{MAE [\%]}&\textbf{DERAEMO1*}&\textbf{DER\_A\_PI*}  \\
        \midrule
        Under& 10.6&1.9\\
        Over&2.3&0.9\\
        \bottomrule
    \end{tabular}
    \label{tab:out_error}
\end{table}
\added{Both models use the parameters optimized in the previous subsection. The results validate the generalization capability of the proposed PI block, as the accuracy closely matches that achieved with the training data in Subsection~\ref{subsec:insample}. Furthermore, DER\_A\_PI again outperforms DERAEMO1 model, particularly during undervoltage events.}
\section{Conclusions and Outlook}
\label{sec:conclusion}

This paper addresses the aggregate modeling of DERs, specifically focusing on their VRT capabilities. DER behavior during voltage disturbances is challenging to predict, and developing accurate and computationally efficient models is essential for the secure operation of future power systems. Performing simulations with detailed DER dynamic models and VRT characteristics may be feasible (although computationally challenging) when modeling one or only a few distribution grids. However, running dynamic simulations for a large-scale power system with many distribution grids quickly becomes computationally prohibitive. In that case, aggregate models become indispensable for power system operational tasks, such as dynamic security assessment or dynamic state estimation. \par
Our results signify that the fractional tripping block of the existing DER\_A model can have limited accuracy during and after large disturbance events. This conclusion is especially relevant if its parameters are not tuned for the specific use case. In this work, we propose an intuitive gray-box model with parameters based on the VRT characteristics of the individual inverter types. The proposed PI fractional tripping model significantly outperforms the existing models in the analyzed scenarios.\par
This study serves as a proof of concept for the proposed modeling technique. As subsequent steps, we will propose a method for tuning the parameters of the proposed model without the need for detailed simulation. This extension could significantly increase the practical applicability of the modeling technique. Furthermore, we envision extending the range of operating conditions and the analyzed disturbance types, and evaluating performance across various distribution grids and DER characteristics.  
\section*{Acknowledgments}
We would like to thank Tom Perrau for his valuable contributions during the project discussions and brainstorming sessions, whose idea developments were instrumental in shaping the direction of this study.
\bibliography{main.bib}
\bibliographystyle{ieeetr}
\appendix
\subsection{Parameter Values of Detailed DER Models}
\label{app:par}
The time delays of Volt--VAr and Volt--Watt control loops significantly limit voltage support during disturbances. These functions were not enabled per default in AS 4777.2-2005, and AS/NZS 4777.2:2015, and the field data suggest their adoption rate was poor \cite{AEMO2019, AEMO_compliance}; therefore, they are not activated in the inverters adherent to these standards. 
\begin{table}[H]
    \centering
    \caption{Parameter values for the detailed DER models.}
    \begin{tabular}{c c c c}
    \toprule
    {\textbf{Parameter}}   & INV2005 & INV2015 & INV2020\\
    \midrule
    $T_{rv}$ [s]  & 0.02& 0.02&0.02 \\ 
       $T_g$ [s]  & 0.02& 0.02&0.02 \\ 
       $T_{iq}$ [s]  & 0.02& 0.02&0.02 \\ 
       $T_\textrm{pord}$ [s]  & 0.02& 0.02&0.02 \\ 
       $K_{qv1}$  & N/A& N/A&7.78 \\ 
       $K_{qv2}$  & N/A& N/A&7.65 \\ 
       $T_{qv}$ [s]  & N/A&N/A&6.0 \\ 
       dbq1  & N/A& N/A&0.956 \\ 
       dbq2  & N/A& N/A&1.043 \\ 
       dbp1  & N/A& N/A&1.07 \\ 
       dbp2  & N/A& N/A&1.1075 \\ 
       $T_{pv}$ [s]  & N/A& N/A&8.0 \\ 
       \bottomrule
    \end{tabular}
    \label{tab:DERD}
\end{table}
The parameters for the fleet of inverters used in the detailed simulation are sampled within the bounds provided by the following table. The corresponding operating regions are shown in Fig.~\ref{fig:VRT}. Momentary cessation deactivation delay is in the range ${[0-0.1]}$ seconds; the reactivation delay is the range $[0-0.4]$ seconds.

\begin{table}[H]
    \centering
    \caption{Bounds on VRT thresholds according to different grid codes.}
    \begin{tabular}{c c c c}
    \toprule
    {\textbf{Parameter}}    & INV2005 & INV2015 & INV2020\\
    \midrule
    $v_{l0}$ [p.u.]  & [0.2--0.8]& [0.1--0.5]&[0.29--0.31] \\ 
       $v_{l1}$ [p.u.]  & [0.84--0.92]& [0.84--0.92]&[0.77--0.79] \\ 
       $v_{h1}$ [p.u.]  & [1.08--1.12]& [1.12--1.14]&[1.14--1.16] \\ 
       $v_{h,mc}$ [p.u.]  & N/A& N/A&[1.12--1.14] \\ 
       $v_{h0}$ [p.u.]  & [1.14--1.20]& [1.14--1.16]&[1.19--1.21] \\ 
       $t_{l0}$[s]  & 0.0& 0.0&[1.0--2.0] \\ 
       $t_{l1}$ [s] & [0.0-0.2]& [0.0--1.0]&10.0 \\ 
       $t_{h1}$ [s]  & [0.0--1.6]&[0.0--1.0]]&[1.0--2.0] \\ 
       $t_{h0}$  [s]& 0.0& 0.0&0.0 \\ 
   
       \bottomrule
    \end{tabular}
    \label{tab:bounds}
\end{table}

\subsection{Parameters of DER\_A and DERAEMO1 fractional tripping blocks}
\begin{table}[H]
    \centering
    \caption{Recommended parameter values for DER tripping logic according to DER\_A \cite{param_DERA} and DERAEMO1 \cite{AEMO_PSSE}.}
    \begin{tabular}{c c c c c}
    \toprule
    {\textbf{Parameter}}     & \multirow{2}{*}{\textbf{DER\_A} }&\multicolumn{3}{c}{\textbf{DERAEMO1} } \\
    {}& {}& 2005 & 2015 & 2020\\
    \midrule
       $v_{l0}$ [p.u.] &0.44  & 0.75& 0.5&0.5 \\ 
       $v_{l1}$ [p.u.] &0.49  & 0.9& 0.9&0.9 \\ 
       $v_{h1}$ [p.u.] &1.15  & 1.13& 1.13&1.19 \\ 
       $v_{h0}$ [p.u.] &1.2  & 1.18& 1.18&1.21 \\ 
       $V_{r, frac}$  &0.35  & 0.625& 0.713&1 \\ 
       $t_{vl0}$ [s] &0.16  & 1.58& 1.77&1.77 \\ 
       $t_{vl1}$ [s] &0.16  & 0.027& 0.037&0.037 \\ 
       $t_{vh0}$ [s] &0.16  & 0.88& 0.16&0.16 \\ 
       $t_{vh1}$ [s] &0.16  & 1.94& 1.87&1.87 \\
       \bottomrule
    \end{tabular}
    
    \label{tab:VRT_DERA}
\end{table}

\subsection{Pseudo Code for the Proposed PI Fractional Tripping Block}
\label{pseudocode}
The block shown in Fig.~\ref{fig:PI_model} is implemented according to the following pseudo-code. The code is executed with a time resolution of $t_s=\SI{0.05}{ms}$. Integrator states $p_{l,\textrm{del}}$ and $p_{l,\textrm{rec}}$ are initialized at $0$, while $v_{\textrm{min}}$ and $v_{\textrm{ss,filt}}$ are initialized at 1~[p.u.]. The code represents undervoltage fractional deactivation; for overvoltage deactivation, the logic is analogous.

\begin{algorithm}
\caption{Pseudo Algorithm for the Proposed PI Model}
\begin{algorithmic}[1]

\State \textbf{Parameters:}
\Statex \hspace{2em} $v_{l0,p}$, $v_{l1,p}$, $v_{l0,i^-}$, $v_{l1,i^-}$, $T_{l,i^-}$, $v_{l0,i^+}$, $T_{l,i^+}$

\State \textbf{Inputs from Previous Step:}
\Statex \hspace{2em} $p_{l,\textrm{del}}$, $p_{l,\textrm{rec}}$, $v_{\textrm{min}}$, $v_\textrm{ss,filt}$

\State \textbf{Inputs at Current Time Step:}
\Statex \hspace{2em} $v_{\textrm{ss}}$ 
        \State \textit{\% Calculate filtered voltage}
        \State $v_{\textrm{ss,filt}} \gets v_{\textrm{ss,filt}} + (v_{\textrm{ss}} - v_{\textrm{ss,filt}}) \times (1 - e^{-t_s/T_{rv}})$

        \State \textit{\% Update minimum voltage}
        
        \State $v_{\textrm{min}} \gets \min (v_{\textrm{ss,filt}}, v_{\textrm{min}})$
        \State \textit{\% Immediate deactivation (proportional term)}
        \State $p_{l,\textrm{im}} \gets (v_{\textrm{min}} - v_{l0,p}) / (v_{l1,p} - v_{l0,p})$
        \State $p_{l,\textrm{im}} \gets \max(0, \min(1, p_{l,\textrm{im}}))$ 

        \State \textit{\% Deactivation integral term}
        \State $p_{{l,\textrm{del\_rate}}} \gets (v_{l1,i^-} - v_{\textrm{ss,filt}}) / (v_{l1,i^-} - v_{l0,i^-})$
        \State $p_{{l,\textrm{del\_rate}}} \gets \max(0, \min(1, p_{{l,\textrm{del\_rate}}}))$
        \State $p_{l,\textrm{del}} \gets p_{l,\textrm{del}} + p_{{l,\textrm{del\_rate}}}  (t_s / T_{l,i^-})$

        \State $p_{{l,\textrm{del\_lim}}} \gets (v_{l1,i^-} - v_{\textrm{min}}) / (v_{l1,i^-} - v_{l0,i^-})$
        \State $p_{{l,\textrm{del\_lim}}} \gets \max(0, \min(1, p_{{l,\textrm{del\_lim}}}))$
        \State $p_{l,\textrm{del}} \gets \min (p_{l,\textrm{del}}, p_{{l,\textrm{del\_lim}}}) $
        \State \textit{\% Reactivation integral term}
        \State $p_{{l,\textrm{rec\_rate}}} \gets (v_{\textrm{ss,filt}} - v_{l0,i^+}) / (v_{l1,i^+} - v_{l0,i^+})$
        \State $p_{{l,\textrm{rec\_rate}}} \gets \max(0, \min(1, p_{{l,\textrm{rec\_rate}}}))$
        \State $p_{l,\textrm{rec}} \gets p_{l,\textrm{rec}} + p_{{l,\textrm{rec\_rate}}}  (t_s / T_{l,i^+}) $

        \State $p_{{l,\textrm{rec\_lim}}} \gets (v_{\textrm{ss,filt}} - v_{l0,i^+}) / (v_{l1,i^+} - v_{l0,i^+})$
        \State $p_{{l,\textrm{rec\_lim}}} \gets \max(0, \min(1, p_{{l,\textrm{rec\_lim}}}))$
       
        \State $p_{\textrm{rec}} \gets \min (p_{l,\textrm{rec}}, p_{{l,\textrm{rec\_lim}}})$
        \State \textit{\% Compute the fraction of active DERs}
        \State $p_{l,\textrm{PI}} \gets p_{l,\textrm{im}} - p_{l,\textrm{del}} + p_{l,\textrm{rec}}$
        \State $p_{l,\textrm{PI}} \gets \max(0, \min(1, p_{l,\textrm{PI}}))$
        \State \textbf{Return:} $p_{l,\textrm{PI}}$, $p_{l,\textrm{del}}$, $p_{l,\textrm{rec}}$, $v_{\textrm{min}}$, $v_\textrm{ss,filt}$

\end{algorithmic}
\end{algorithm}

\end{document}